# Lossless compression catalyst based on binary allocation via modular arithmetic

**Mario Mastriani**

DLQS LLC, 4431 NW 63RD Drive, Coconut Creek, FL 33073, USA.

mmastri@gmail.com

***Abstract***— A new binary (bit-level) lossless compression catalyst method based on a modular arithmetic, called Binary Allocation via Modular Arithmetic (BAMA), has been introduced in this paper. In other words, BAMA is for storage and transmission of binary sequences, digital signal, images and video, also streaming and all kinds of digital transmission. As we said, our method does not compress, but facilitates the action of the real compressor, in our case, any lossless compression algorithm (Run Length Encoding, Lempel-Ziv-Welch, Huffman, Arithmetic, etc), that is, it acts as a compression catalyst. Finally, this catalyst allows a significant increase in the compression performance of binary sequences, among others.

## I. INTRODUCTION

Data compression is the process of converting an input data stream (the source stream or the original raw data) into another data stream (the output, the bitstream, or the compressed stream) that has a smaller size [1]. A stream is either a file or a buffer in memory. Data compression is popular for two reasons: (1) People like to accumulate data and hate to throw anything away. No matter how big a storage device one has, sooner or later it is going to overflow. Data compression seems useful because it delays this inevitability. (2) People hate to wait a long time for data transfers. When sitting at the computer, waiting for a Web page to come in or for a file to download, we naturally feel that anything longer than a few seconds is a long time to wait. The field of data compression is often called source coding. We imagine that the input symbols (such as bits, ASCII codes, bytes, audio samples, or pixel values) are emitted by a certain information source and have to be coded before being sent to their destination. The source can be memoryless, or it can have memory. In the former case, each symbol is independent of its predecessors. In the latter case, each symbol depends on some of its predecessors and, perhaps, also on its successors, so they are correlated. A memoryless source is also termed "independent and identically distributed" or IIID. Data compression has come of age in the last 20 years. Both the quantity and the quality of the body of literature in this field provide ample proof of this.

There are many known methods for data compression. They are based on different ideas, are suitable for different types of data, and produce different results, but they are all based on the same principle, namely they compress data by removing redundancy from the original data in the source file. Any nonrandom data has some structure, and this structure can be exploited to achieve a smaller representation of the data, a representation where no structure is discernible. The terms redundancy and structure are used in the professional literature, as well as smoothness, coherence, and correlation; they all refer to the same thing. Thus, redundancy is a key concept in any discussion of data compression.

The idea of compression by reducing redundancy suggests the general law of data compression, which is to "assign short codes to common events (symbols or phrases) and long codes to rare events." There are many ways to implement this law, and an analysis of any compression method shows that, deep inside, it works by obeying the general law. In computer science and information theory, data compression, source coding, or bit-rate reduction involves encoding information using fewer bits than the original representation.

Compressing data is done by changing its representation from inefficient (i.e., long) to efficient (short). Compression is therefore possible only because data is normally represented in the computer in a format that is longer than absolutely necessary. The reason that inefficient (long) data representations are used all the time is that they make it easier to process the data, and data processing is more common and more important than data compression. The ASCII code for characters is a good example of a data representation that is longer than absolutely necessary. It uses 7-bit codes because fixed-size codes are easy to work with. A variable-size code, however, would be more efficient, since certain characters are used more than others and so could be assigned shorter codes.

In a world where data is always represented by its shortest possible format, there would therefore be no way to compress data. Instead of writing books on data compression, authors in such a world would write books on how to determine the shortest format for different types of data.

The principle of compressing by removing redundancy also answers the following question: "Why is it that an already compressed file cannot be compressed further?" The answer, of course, is that such a file has little or no redundancy, so there is nothing to remove. An example of such a file is random text. In such text, each letter occurs with equal probability, so assigning them fixed-size codes does not add any redundancy. When such a file is compressed, there is no redundancy to remove. (Another answer is that if it were possible to compress an already compressed file, then successive compressions would reduce the size of the file until it becomes a single byte, or even a single bit. This, of course, is ridiculous since a single byte cannot contain the information present in an arbitrarily large file.)

It is therefore clear that no compression method can hope to compress all files or even a significant percentage of them. In order to compress a data file, the compression algorithm has to examine the data, find redundancies in it, and try to remove them. The redundancies in data depend on the type of data (text, images, sound, etc.), which is why a new compression method has to be developed for a specific type of data and it performs best on this type. There is no such thing as a universal, efficient data compression algorithm in the practice.

Data compression has become so important that some researchers (see, for example, [2, 3]) have proposed the SP theory (for "simplicity" and "power"), which suggests that all computing is compression! Specifically, it says: Data compression may be interpreted as a process of removing unnecessary complexity (redundancy) in information, and thereby maximizing simplicity while preserving as much as possible of its nonredundant descriptive power. SP theory is based on the following conjectures:

- All kinds of computing and formal reasoning may usefully be understood as information compression by pattern matching, unification, and search.

- The process of finding redundancy and removing it may always be understood at a fundamental level as a process of searching for patterns that match each other, and merging or unifying repeated instances of any pattern to make one.

In [1], author discusses many compression methods, some suitable for text and others for graphical data (still images or movies) or for audio. Most methods are classified into four categories: run length encoding (RLE), statistical methods, dictionary-based (sometimes called LZ, that is to say, LZ77, LZ78 and LZW) methods, and transforms.

Before delving into the details, we discuss important data compression terms.

- The compressor or encoder is the program that compresses the raw data in the input stream and creates an output stream with compressed (low-redundancy) data. The decompressor or decoder converts in the opposite direction. Note that the term encoding is very general and has several meanings, but since we discuss only data compression, we use the name encoder to mean data compressor. The term codec is sometimes used to describe both the encoder and the decoder. Similarly, the term companding is short for "compressing/expanding."

- Although the literature [1], the term “stream” is used instead of file, stream is a more general term because the compressed data may be transmitted directly to the decoder, instead of being written to a file and saved. Also, the data to be compressed may be downloaded from a network instead of being input from a file.

- For the original input stream, we use the terms unencoded, raw, or original data. The contents of the final, compressed, stream are considered the encoded or compressed data. The term bitstream is also used in the literature to indicate the compressed stream.

- A ***nonadaptive*** compression method is rigid and does not modify its operations, its parameters, or its tables in response to the particular data being compressed. Such a method is best used to compress data that is all of a single type. Examples are the Group 3 and Group 4 methods for facsimile compression [1]. They are specifically designed for facsimile compression and would do a poor job compressing any other data. In contrast, an adaptive method examines the raw data and modifies its operations and/or its parameters accordingly. An example is the adaptive Huffman method [1]. Some compression methods use a 2-pass algorithm, where the first pass reads the input stream to collect statistics on the data to be compressed, and the second pass does the actual compressing using parameters set by the first pass. Such a method may be called semiadaptive. A data compression method can also be locally adaptive, meaning it adapts itself to local conditions in the input stream and varies this adaptation as it moves from area to area in the input. An example is the move-to-front method.

- ***Lossy/lossless compression:*** Certain compression methods are lossy. They achieve better compression by losing some information. When the compressed stream is decompressed, the result is not identical to the original data stream. Such a method makes sense especially in compressing images, movies, or sounds. If the loss of data is small, we may not be able to tell the difference. In contrast, text files, especially files containing computer programs, may become worthless if even one bit gets modified. Such files should be compressed only by a lossless compression method. [Two points should be mentioned regarding text files: (1) If a text file contains the source code of a program, consecutive blank spaces can often be replaced by a single space. (2) When the output of a word processor is saved in a text file, the file may contain information about the different fonts used in the text. Such information may be discarded if the user is interested in saving just the text.]

- ***Cascaded compression:*** The difference between lossless and lossy codecs can be illuminated by considering a cascade of compressions. Imagine a data file A that has been compressed by an encoder X, resulting in a compressed file B. It is possible, although pointless, to pass B through another encoder Y , to produce a third compressed file C. The point is that if methods X and Y are lossless, then decoding C by Y will produce an exact B, which when decoded by X will yield the original file A. However, if any of the compression algorithms is lossy, then decoding C by Y may produce a file B' different from B. Passing B' through X may produce something very different from A and may also result in an error, because X may not be able to read B'.

- ***Perceptive compression:*** A lossy encoder must take advantage of the special type of data being compressed. It should delete only data whose absence would not be detected by our senses. Such an encoder must therefore employ algorithms based on our understanding of psychoacoustic and psychovisual perception, so it is often referred to as a perceptive encoder. Such an encoder can be made to operate at a constant compression ratio, where for each x bits of raw data, it outputs y bits of compressed data. This is convenient in cases where the compressed stream has to be transmitted at a constant rate. The trade-off is a variable subjective quality. Parts of the original data that are difficult to compress may, after decompression, look (or sound) bad. Such parts may require more than y bits of output for x bits of input.

- ***Symmetrical compression*** is the case where the compressor and decompressor use basically the same algorithm but work in “opposite” directions. Such a method makes sense for general work, where the same number of files is compressed as is decompressed. In an asymmetric compression method, either the compressor or the decompressor may have to work significantly harder. Such methods have their uses and are not necessarily bad. A compression method where the compressor executes a slow, complex algorithm and the decompressor is simple is a natural choice when files are compressed into an archive, where they will be decompressed and used very often. The opposite case is useful in environments where files are updated all the time and backups are made. There is a small chance that a backup file will be used, so the decompressor isn't used very often.

- Many modern compression methods are asymmetric. Often, the formal description (the standard) of such a method consists of the decoder and the format of the compressed stream, but does not discuss the operation of the encoder. Any encoder that generates a correct compressed stream is considered ***compliant***, as is also any decoder that can read and decode such a stream. The advantage of such a description is that anyone is free to develop and implement new, sophisticated algorithms for the encoder. The implementor need not even publish the details of the encoder and may consider it proprietary. If a compliant encoder is demonstrably better than competing encoders, it may become a commercial success. In such a scheme, the encoder is considered algorithmic, while the decoder, which is normally much simpler, is termed ***deterministic***. A good example of this approach is the MPEG-1 audio compression method [1].

- A data compression method is called ***universal*** if the compressor and decompressor do not know the statistics of the input stream. A universal method is ***optimal*** if the compressor can produce compression factors that asymptotically approach the entropy of the input stream for long inputs.

- The term ***file differencing*** refers to any method that locates and compresses the differences between two files. Imagine a file A with two copies that are kept by two users. When a copy is updated by one user, it should be sent to the other user, to keep the two copies identical. Instead of sending a copy of A, which may be big, a much smaller file containing just the differences, in compressed format, can be sent and used at the receiving end to update the copy of A. In [1], the author discusses some of the details and shows why compression can be considered a special case of file differencing. Note that the term "differencing" is used in [1] to describe a completely different compression method.

- Most compression methods operate in the ***streaming mode***, where the codec inputs a byte or several bytes, processes them, and continues until an end-of-file is sensed. Some methods, such as Burrows-Wheeler [1], work in the block mode, where the input stream is read block by block and each block is encoded separately. The block size in this case should be a user-controlled parameter, since its size may greatly affect the performance of the method.

- Most compression methods ***are physical***. They look only at the bits in the input stream and ignore the meaning of the data items in the input (e.g., the data items may be words, pixels, or audio samples). Such a method translates one bit stream into another, shorter, one. The only way to make sense of the output stream (to decode it) is by knowing how it was encoded. Some compression methods are logical. They look at individual data items in the source stream and replace common items with short codes. Such a method is normally special purpose and can be used successfully on certain types of data only. The pattern substitution method described in [1] is an example of a logical compression method.

Moreover with dynamic dictionaries the efficiency of the compression converges slowly with the size of the data [4]. For this reason, specialized decorrelation transforms have been successively studied with the aim at further improving the compression ratio [5]. Among existing decorrelation transforms, fixed-basis (heuristic) transforms, such as discrete cosine transform, are used in practice while variable-basis (optimal) transforms, such as KLT [5-7], are used to bound the best theoretical transform performance. Indeed, the variability of the transform basis is not a desirable property in data compression. Fixed-basis transforms usually target a specific class of data, for example the discrete cosine transform is advantageous for imaging (JPEG standard [8, 9]), similar case to discrete wavelet transform (JPEG2000 standard [10]).

Finally, a lossless compression scheme refers actually to two algorithms: (i) a compression algorithm that takes in input a data sequence X and reduces it to Xc (that requires fewer, or at most equal, bits than X), and (ii) a reconstruction algorithm that recovers exactly X from Xc, i.e., with no loss of information. In other words, in a lossless compression scheme the compression algorithm must be reversible [5]. Given the dual nature of lossless compression schemes, we concentrate hereafter only on compression algorithms provided that they satisfy the reversibility requirement. Nowadays, practical compression approaches are based on two subsequent phases: first, the input data is decorrelated and then entropy encoding techniques are applied as final compression step.

1. ***Data Decorrelation (Entropy reduction):*** Data decorrelation, also called entropy reduction, is a data preprocessing phase aiming at reducing the autocorrelation of the input data. Linear transformations are efficient means to accomplish this task [5]. Among all the decorrelation transforms, the best coding gain, in term of compression ratio, is provided by the Karhunen-Loeve Transform (KLT). The KLT is a transform having signal-dependent basis and random variables as coefficients. In the discrete (binary) domain, the KLT is described by a matrix having as columns the eigenvectors of the autocorrelation matrix of the input (binary) sequence considered [5]. The signal-dependency of the KLT makes its use inefficient in practical applications where the basis (or the data autocorrelation itself) must be provided to the decoding algorithm which is unaware of the properties of the compressed data. Such overhead removes the advantage of the KLT theoretical optimality. For this reason, the KLT is typically used as theoretical bound for the best coding gain achievable by decorrelation transforms. On the other hand, practical decorrelation techniques of interest make assumptions on the original data properties resulting in fixed basis transforms. A well-known example is the Discrete Cosine Transform (DCT), widely used in image compression methods, e.g., JPEG [8, 9], which employs cosine functions as fixed transform basis, and JPEG2000 [10], which employs wavelets as fixed transform basis. Other specialized transforms have been developed in literature, e.g., BCJ/BCJ-2 [11] for binary executables, Burrows-Wheeler [12] especially efficient in the field of bio-informatics etc. Another notable approach to decorrelate data is based on dictionary techniques. The core idea of dictionary techniques  is to build a dictionary (static or dynamic) of recurring patterns in the input data. Such patterns are directly encoded by their indexes in the dictionary. The Lempel-Ziv (LZ) method [13] is one of the most recognized dictionary based compression technique used in many practical compression standards. The general efficiency of existing dictionary techniques is limited by (i) assumptions on pattern recurrence locality and (ii) issues related to dictionary flexibility. We refer the interested reader to [5] for an extended and complete discussion of data decorrelation.

2. ***Entropy Coding:*** Entropy encoding refers to a class of (lossless) coding techniques able to compress an input data down to its entropy. When the entropy information is defined according to the exact probabilistic model, entropy encoding achieves the optimum compression for any input data. However, the correlation and the (often complex) underlying function that produced the data set is typically not known. Hence, the choice of the right probabilistic model is a difficult problem usually simplified by decorrelating transforms. Once the data is (fully) decorrelated, a simple stochastic model is reliably employable to define the information entropy. Then, entropy encoding methods can be applied successfully. Original entropy encoding techniques are Huffman coding [14] and arithmetic coding [5] that form the basis of current compression software and standards. For a review of such methods, we refer again to [5].

In this paper we propose A new binary (bit-level) lossless compression catalyst method based on a modular arithmetic, called Binary Allocation via Modular Arithmetic (BAMA), for storage and transmission of binary sequences, digital signal, images and video, also streaming and all kinds of digital transmission. This method does not compress, but facilitates the action of the real compressor, in our case, any lossless compression algorithm (Run Length Encoding, Lempel-Ziv-Welch, Huffman, Arithmetic, etc), that is, it acts as a catalyst compression, allowing a significant increase in the compression performance of binary sequences, among others. In fact, this technology is ideal in all cases where the volume of information to be stored is extravagantly huge. Moreover, since its implementation is extremely simple (low computational cost), this constitutes a formidable tool for storage systems related to social networs.

The Modular Arithmetic (in general) and the congruence concept (in particular) are outlined in Section II. BAMA is presented in Section III. The benchmarking performance is outlined in Section IV. In Section V, we discuss briefly the possible applications. Finally, Section VI provides a conclusion of the paper.

## II. Modular Arithmetic

### a) Integer Arithmetic:

In integer arithmetic, we use a set and a few operations. You are familiar with this set and the correspond-ding operations, but they are reviewed here to create a background for modular arithmetic [15-27].

***Set of Integers***

The set of integers, denoted by Z, contains all integral numbers (with no fraction) from negative infinity to positive infinity, see Fig.1.

$$\mathbf{Z} = \{\ldots, -2, -1, 0, 1, 2, \ldots\}$$

Fig.1: The set of integers.

***Binary Operations***

In cryptography, we are interested in three binary operations applied to the set of integers. A binary operation takes two inputs and creates one output. Three common binary operations defined for integers are *addition*, *subtraction*, and *multiplication*. Each of these operations takes two inputs (*a* and *b*) and creates one output (*c*) as shown in Fig.2. The two inputs come from the set of integers; the output goes into the set of integers. Note that *division* does not fit in this category because, as we will see shortly, it produces two outputs instead of one.

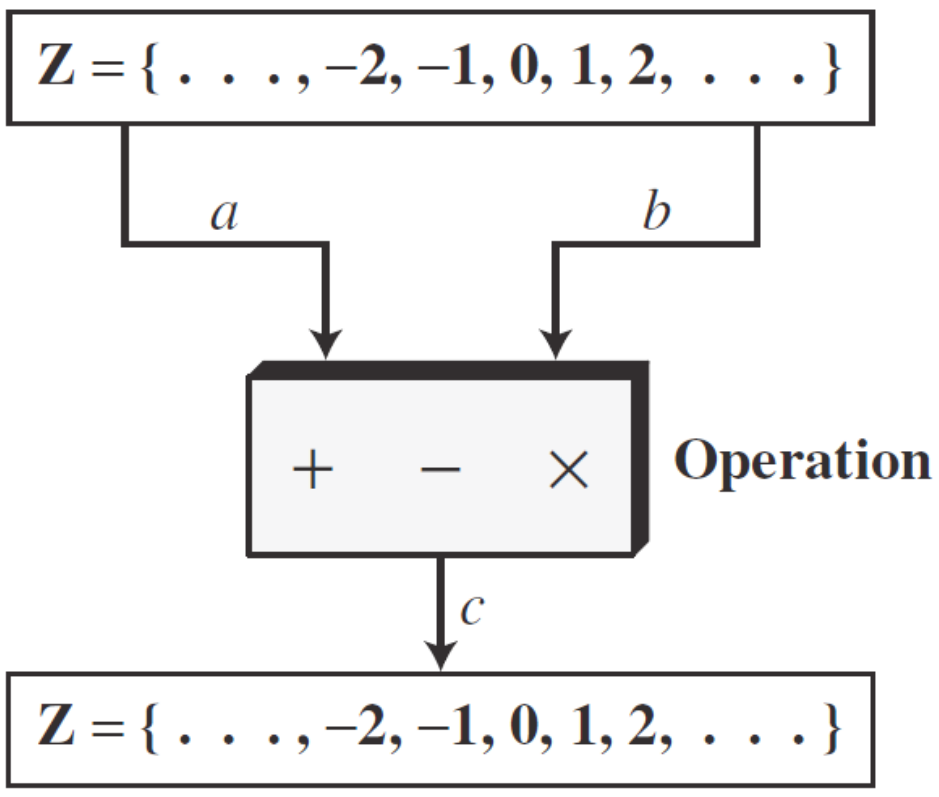

Fig.2: Three binary operations for the set of integers.

The following shows the results of the three binary operations on two integers. Because each input can be either positive or negative, we can have four cases for each operation.

| | | | | |
|---|---|---|---|---|
| Add: | $5 + 9 = 14$ | $(-5) + 9 = 4$ | $5 + (-9) = -4$ | $(-5) + (-9) = -14$ |
| Subtract: | $5 - 9 = -4$ | $(-5) - 9 = -14$ | $5 - (-9) = 14$ | $(-5) - (-9) = +4$ |
| Multiply: | $5 \times 9 = 45$ | $(-5) \times 9 = -45$ | $5 \times (-9) = -45$ | $(-5) \times (-9) = 45$ |

***Integer Division***

In integer arithmetic, if we divide *p* by *n*, we can get *a* and *b*. The relationship between these four integers can be shown as

$$p = a \times n + b \qquad (1)$$

In this relation, $p$ is called the *dividend; a*, the *quotient; n*, the *divisor;* and $b$, the *remainder*. Note that this is not an operation, because the result of dividing $p$ by $n$ is two integers, $a$ and $b$. We can call it *division relation*. For example, assume that $p$ = 255 and $n$ = 11. We can find $a$ = 23 and $b$ = 2 using the division algorithm we have learned in arithmetic as shown in Fig.3.

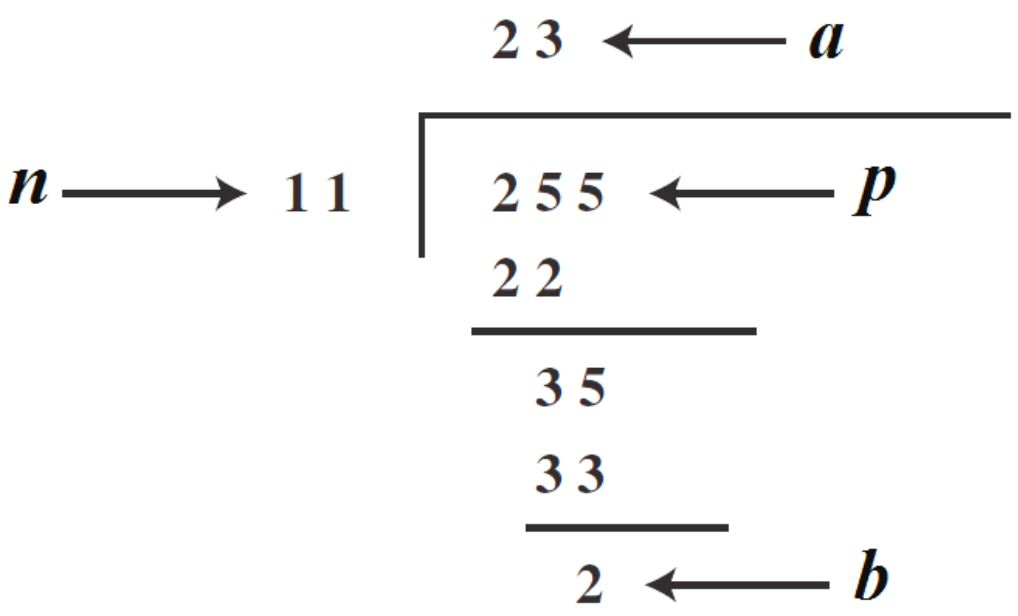

Fig.3: Finding the quotient and the remainder.

Most computer languages can find the quotient and the remainder using language specific operators. For example, in the MATLAB® [28] language, the operator / can find the quotient, and the operator % can find the remainder.

***Two Restrictions***

When we use the above division relationship in cryptography, we impose two restrictions. First, we require that the divisor be a positive integer ($n > 0$). Second, we require that the remainder be a nonnegative integer ($b \geq 0$). Fig.4 shows this relationship with the two above-mentioned restrictions.

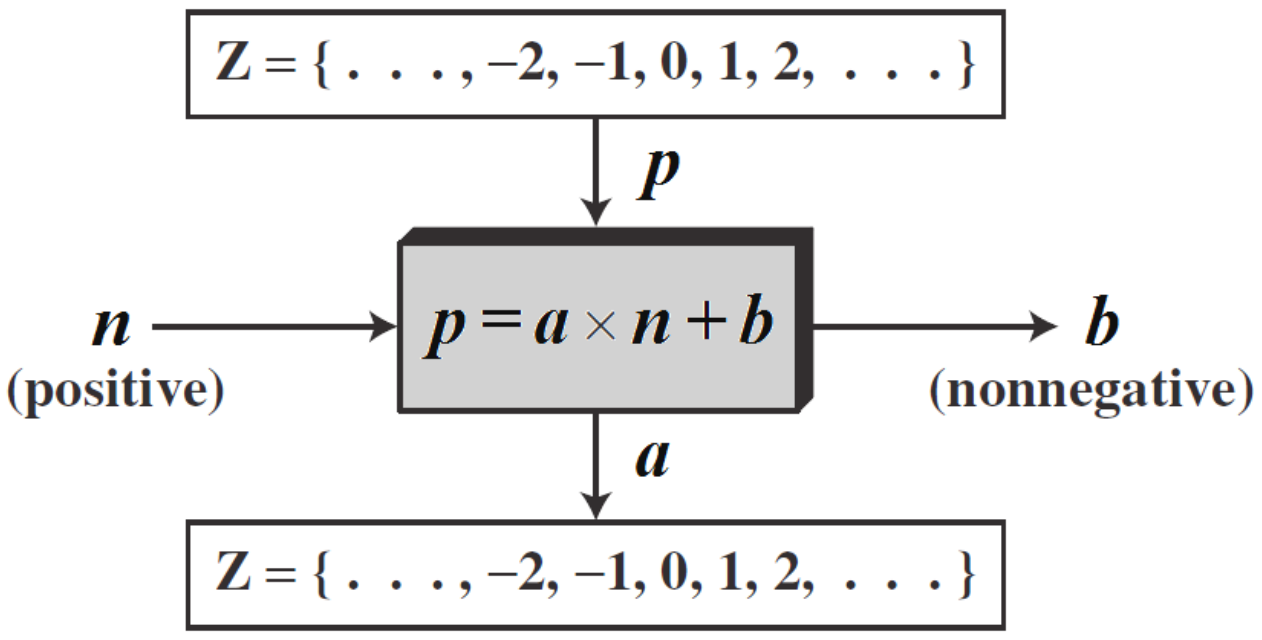

Fig.4: Division algorithm for integers.

***The Graph of the Relation***

We can show the above relation with the two restrictions on $n$ and $b$ using two graphs in Fig.5. The first one shows the case when $p$ is positive; the second when $p$ is negative.

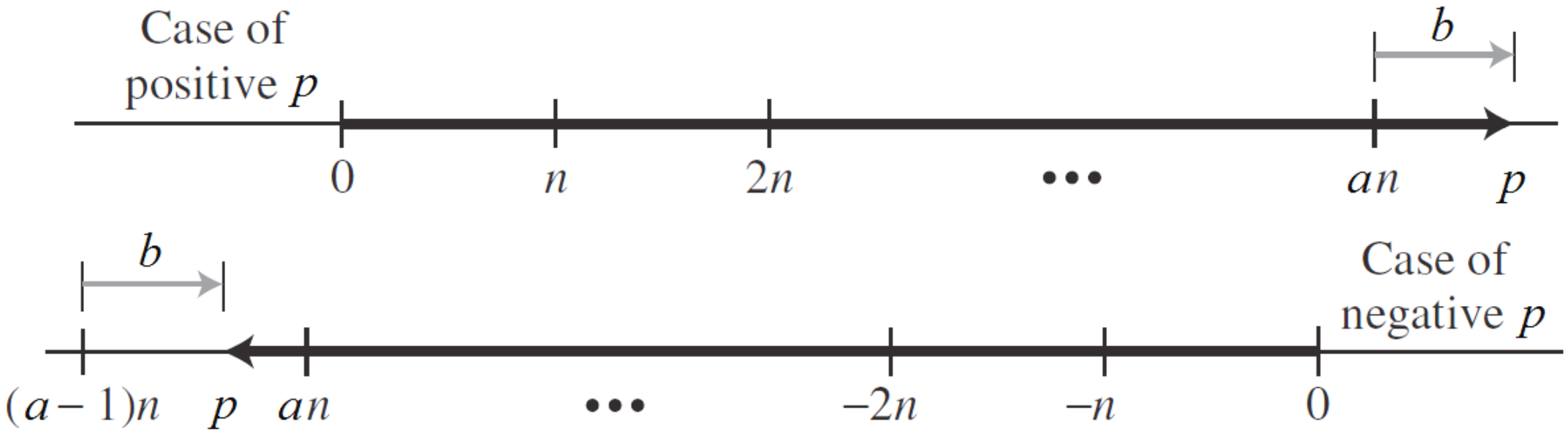

Fig.5: Graph of division algorithm.

Starting from zero, the graph shows how we can reach the point representing the integer $p$ on the line. In case of a positive $p$, we need to move $a \times n$ units to the right and then move extra $b$ units in the same direction. In case of a negative $p$, we need to move $(a - 1) \times n$ units to the left ($a$ is negative in this case) and then move $b$ units in the opposite direction. In both cases the value of $b$ is positive.

***Divisibility***

Let us briefly discuss divisibility, a topic we often encounter in cryptography. If $p$ is not zero and we let $b = 0$ in the division relation, we get

$$p = a \times n \qquad (2)$$

## b) Modular Arithmetic:

The division relationship ($p = a \times n + b$) discussed in the previous section has two inputs ($p$ and $n$) and two outputs ($a$ and $b$). In modular arithmetic, we are interested in only one of the outputs, the remainder $b$. We don't care about the quotient $a$. In other words, we want to know what is the value of $b$ when we divide $p$ by $n$. This implies that we can change the above relation into a binary operator with two inputs $p$ and $n$ and one output $b$.

**Modulo Operator**

The above-mentioned binary operator is called the modula operator and is shown as *mod*. The second input ($n$) is called the modulus. The output $b$ is called the residue. Figure 6 shows the division relation compared with the modulo operator.

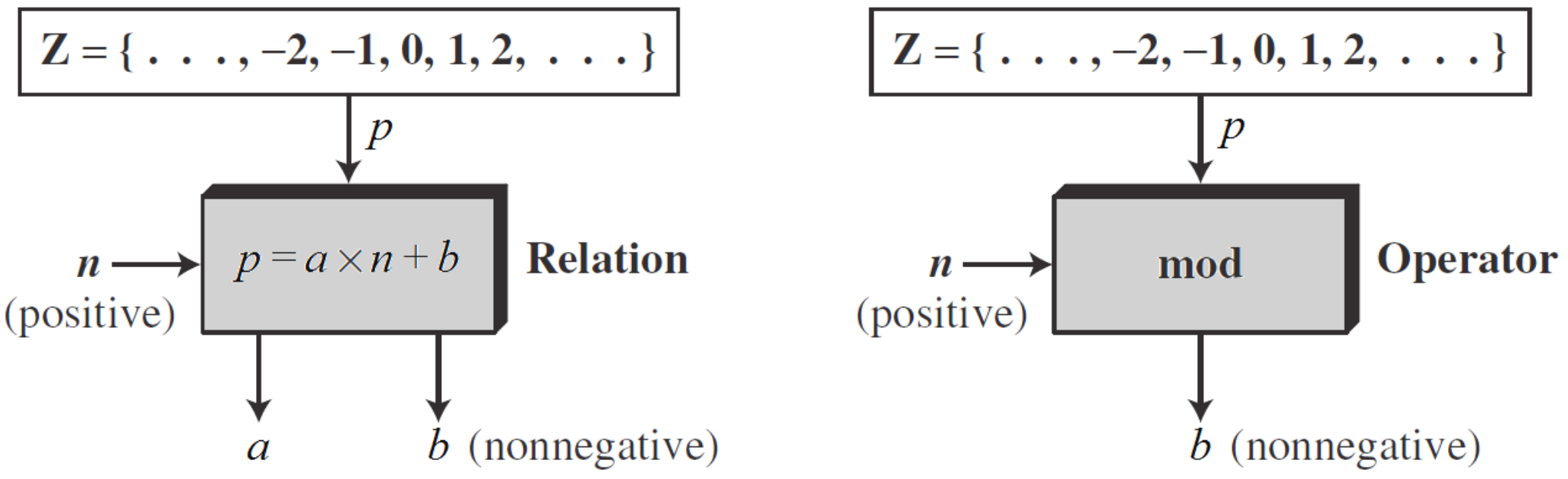

Fig.6: Division relation and modulo operator.

As Fig.6 shows, the modula operator (**mod**) takes an integer ($p$) from the set **Z** and a positive modulus ($n$). The operator creates a nonnegative residue ($b$). We can say.

$$p \bmod n = b \qquad (3)$$

**Set of Residues: $\mathbf{Z}_n$**

The result of the modula operation with modulus $n$ is always an integer between 0 and $n - 1$. In other words, the result of $p$ mod $n$ is always a nonnegative integer less than $n$. We can say that the modula operation creates a set, which in modular arithmetic is referred to as the **set of least residues modulo *n***, or $\mathbf{Z}_n$.

$$\mathbf{Z}_n = \{ 0, 1, 2, 3, \ldots, (n - 1) \}$$

$$\mathbf{Z}_2 = \{ 0, 1 \} \quad \mathbf{Z}_6 = \{ 0, 1, 2, 3, 4, 5 \} \quad \mathbf{Z}_{11} = \{ 0, 1, 2, 3, 4, 5, 6, 7, 8, 9, 10 \}$$

Fig.7: Some $\mathbf{Z}_n$ sets.

However, we need to remember that although we have only one set of integers (**Z**), we have infinite instances of the set of residues ($\mathbf{Z}_n$), one for each value of $n$. Figure 7 shows the set $\mathbf{Z}_n$ and three instances, $\mathbf{Z}_2$, $\mathbf{Z}_6$, and $\mathbf{Z}_{11}$.

**Congruence**

In cryptography [29-35], we often used the concept of **congruence** instead of equality. Mapping from **Z** to $\mathbf{Z}_n$ is not one-to-one. Infinite members of **Z** can map to one member of $\mathbf{Z}_n$. For example, the result of 2 mod 10 = 2, 12 mod 10 = 2, 22 mod 2 = 2, and so on. In modular arithmetic, integers like 2, 12, and 22 are called congruent mod 10. To show that two integers are congruent, we use the **congruence operator** (≡). We add the phrase (mod $n$) to the right side of the congruence to define the value of modulus that makes the relationship valid. For example, we write:

| | | | |
|---|---|---|---|
| $2 \equiv 12 \pmod{10}$ | $13 \equiv 23 \pmod{10}$ | $34 \equiv 24 \pmod{10}$ | $-8 \equiv 12 \pmod{10}$ |
| $3 \equiv 8 \pmod{5}$ | $8 \equiv 13 \pmod{5}$ | $23 \equiv 33 \pmod{5}$ | $-8 \equiv 2 \pmod{5}$ |

Figure 8 shows the idea of congruence. We need to explain several points, for the purpose of being able to understand the main idea behind this theory. Such points are:

a. The congruence operator looks like the equality operator, but there are differences. First, an equality operator maps a member of **Z** to itself; the congruence operator maps a member from **Z** to a member of $\mathbf{Z}_n$. Second, the equality operator is one-to-one; the congruence operator is many-to-one.

b. The phrase (mod $n$) that we insert at the right-hand side of the congruence operator is just an indication of the destination set ($\mathbf{Z}_n$). We need to add this phrase to show what modulus is used in the mapping. The symbol *mod* used here does not have the same meaning as the binary operator. In other words, the symbol *mod* in 12 mod 10 is an operator; the phrase (mod 10) in 2 α 12 (mod 10) means that the destination set is $\mathbf{Z}_{10}$.

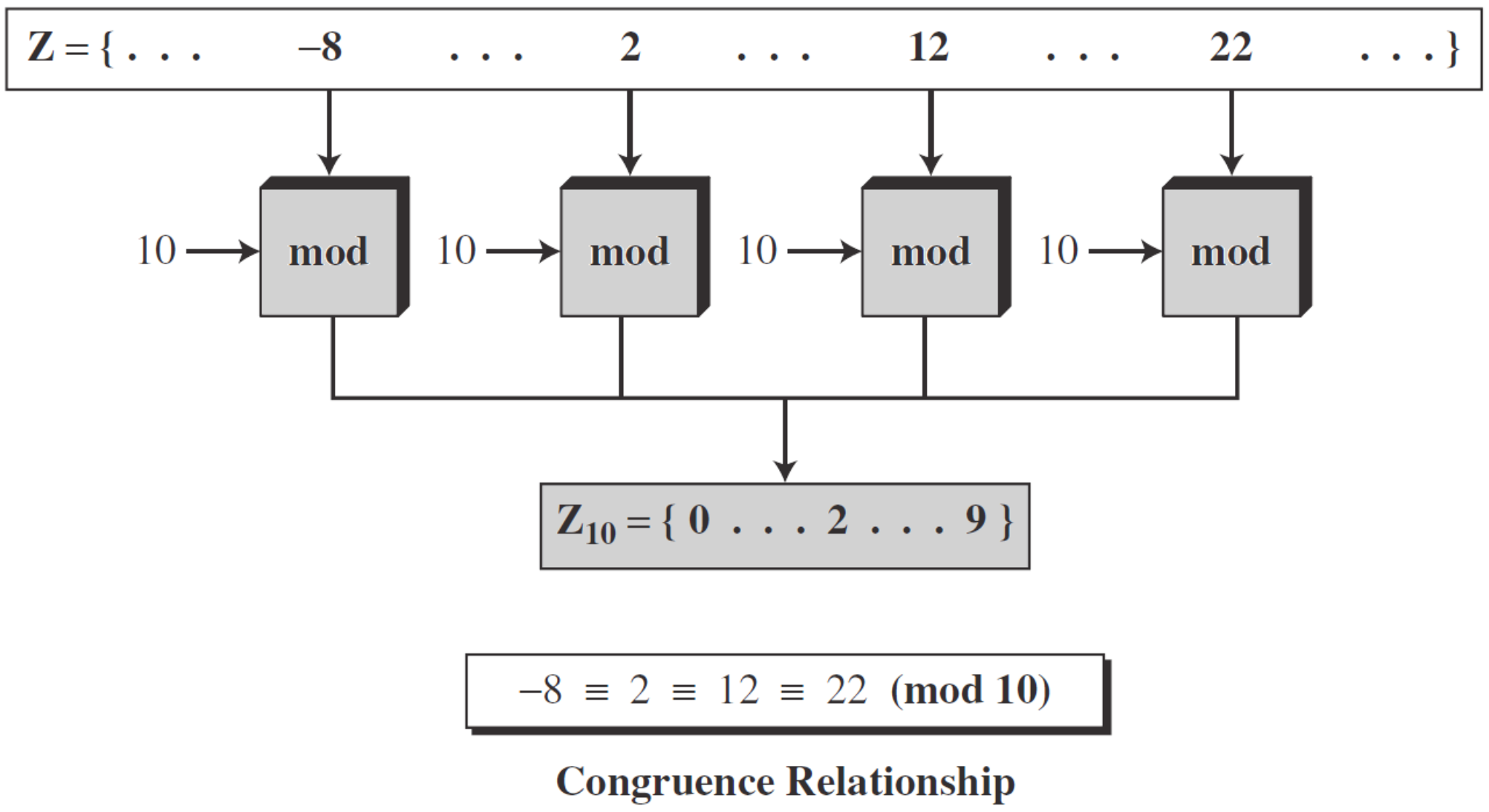

Fig.8: Concept of congruence.

***Residue Classes***

A **residue class** ]$p$[ or ]$p$[$_n$ is the set of integers congruent modulo $n$. In other words, it is the set of all integers such that $x = p \pmod{n}$. For example, if $n = 5$, we have five sets ]0[, ]1[, ]2[, ]3[, and ]4[ as shown below:

$$
\begin{aligned}
]0[ &= \{\ldots, -15, -10, -5, 0, 5, 10, 15, \ldots\} \\
]1[ &= \{\ldots, -14, -9, -4, 1, 6, 11, 16, \ldots\} \\
]2[ &= \{\ldots, -13, -8, -3, 2, 7, 12, 17, \ldots\} \\
]3[ &= \{\ldots, -12, -7, -5, 3, 8, 13, 18, \ldots\} \\
]4[ &= \{\ldots, -11, -6, -1, 4, 9, 14, 19, \ldots\}
\end{aligned}
$$

The integers in the set ]0[ are all reduced to 0 when we apply the modulo 5 operation on them. The integers in the set ]1[ are all reduced to 1 when we apply the modulo 5 operation, and so on. In each set, there is one element called the least (nonnegative) residue. In the set ]0[, this element is 0; in the set ]1[, this element is 1; and so on. The set of all of these least residues is what we have shown as $\mathbf{Z}_5 = \{0, 1, 2, 3, 4\}$. In other words, the set $\mathbf{Z}_n$ is the set of all **least residue** modulo $n$.

***Circular Notation***

The concept of congruence can be better understood with the use of a circle. Just as we use a line to show the distribution of integers in **Z**, we can use a circle to show the distribution of integers in $\mathbf{Z}_n$. Figure 9 shows the comparison between the two. Integers 0 to $n - 1$ are spaced evenly around a circle. All congruent integers modulo $n$ occupy the same point on the circle. Positive and negative integers from **Z** are mapped to the circle in such a way that there is a symmetry between them.

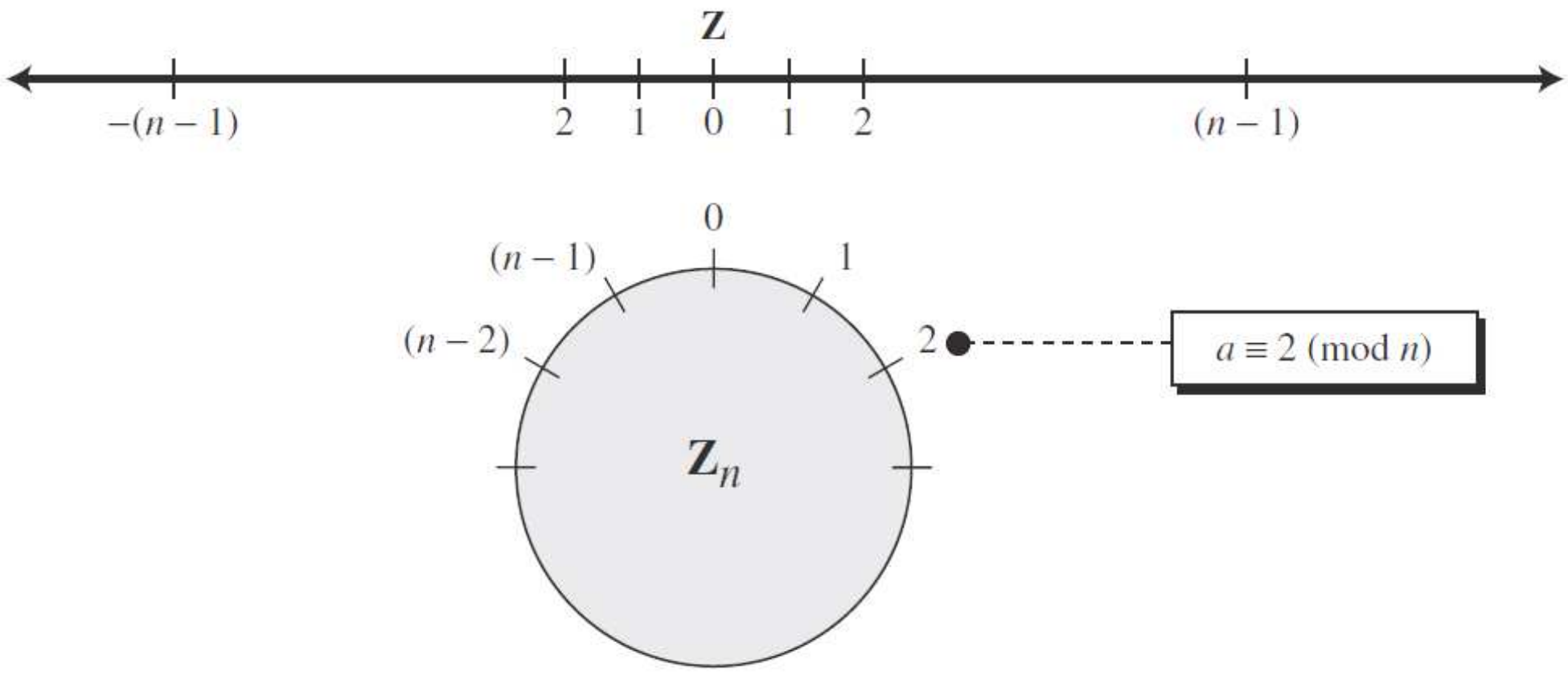

Fig.9: Comparison of **Z** and $\mathbf{Z}_n$ using graphs.

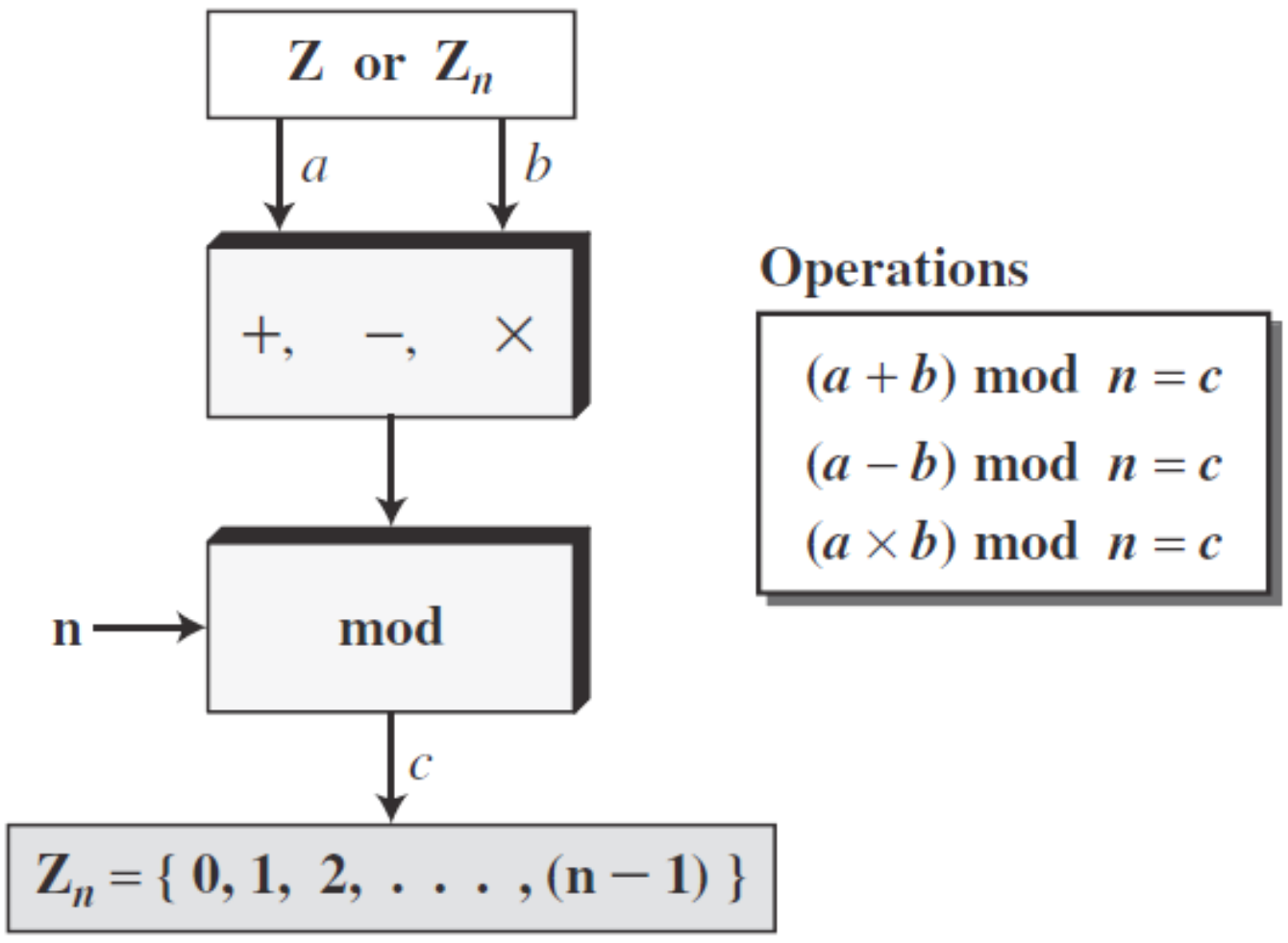

Fig.10: Binary operations in $\mathbf{Z}_n$.

**Operations in $\mathbf{Z}_n$**

The three binary operations (*addition*, *subtraction*, and *multiplication*) that we discussed for the set **Z** can also be defined for the set $\mathbf{Z}_n$. The result may need to be mapped to $\mathbf{Z}_n$ using the mod operator as shown in Fig.10.

Actually, two sets of operators are used here. The first set is one of the binary operators (+, −, ×); the second is the mod operator. We need to use parentheses to emphasize the order of operations. As Figure 10 shows, the inputs ($a$ and $b$) can be members of $\mathbf{Z}_n$ or **Z**.

***Properties***

We mentioned that the two inputs to the three binary operations in the modular arithmetic can come from **Z** or $\mathbf{Z}_n$. The following properties allow us to first map the two inputs to $\mathbf{Z}_n$ (if they are coming from **Z**) before applying the three binary operations (+, −, ×). Interested readers can find proofs for these properties in [15].

| | |
|---|---|
| **First Property:** | $(a + b) \bmod n = [(a \bmod n) + (b \bmod n)] \bmod n$ |
| **Second Property:** | $(a - b) \bmod n = [(a \bmod n) - (b \bmod n)] \bmod n$ |
| **Third Property:** | $(a \times b) \bmod n = [(a \bmod n) \times (b \bmod n)] \bmod n$ |

Figure 11 shows the process before and after applying the above properties. Although the figure shows that the process is longer if we apply the above properties, we should remember that in cryptography we are dealing with very large integers. For example, if we multiply a very large integer by another very large integer, we may have an integer that is too large to be stored in the computer. Applying the above properties make the first two operands smaller before the multiplication operation is applied. In other words, the properties us with smaller numbers. This fact will manifest itself more clearly in discussion of the exponential operation in later chapters.

Above, we have described the main properties of Modular Arithmetic necessaries to the understanding of BAMA. Although these tools are the mathematical basis of modern Cryptography [29-35], so is the development of this lossless compression catalyst called BAMA, which allows a significant increase in the compression performance of binary sequences [36-38] and binary string coding [39-42].

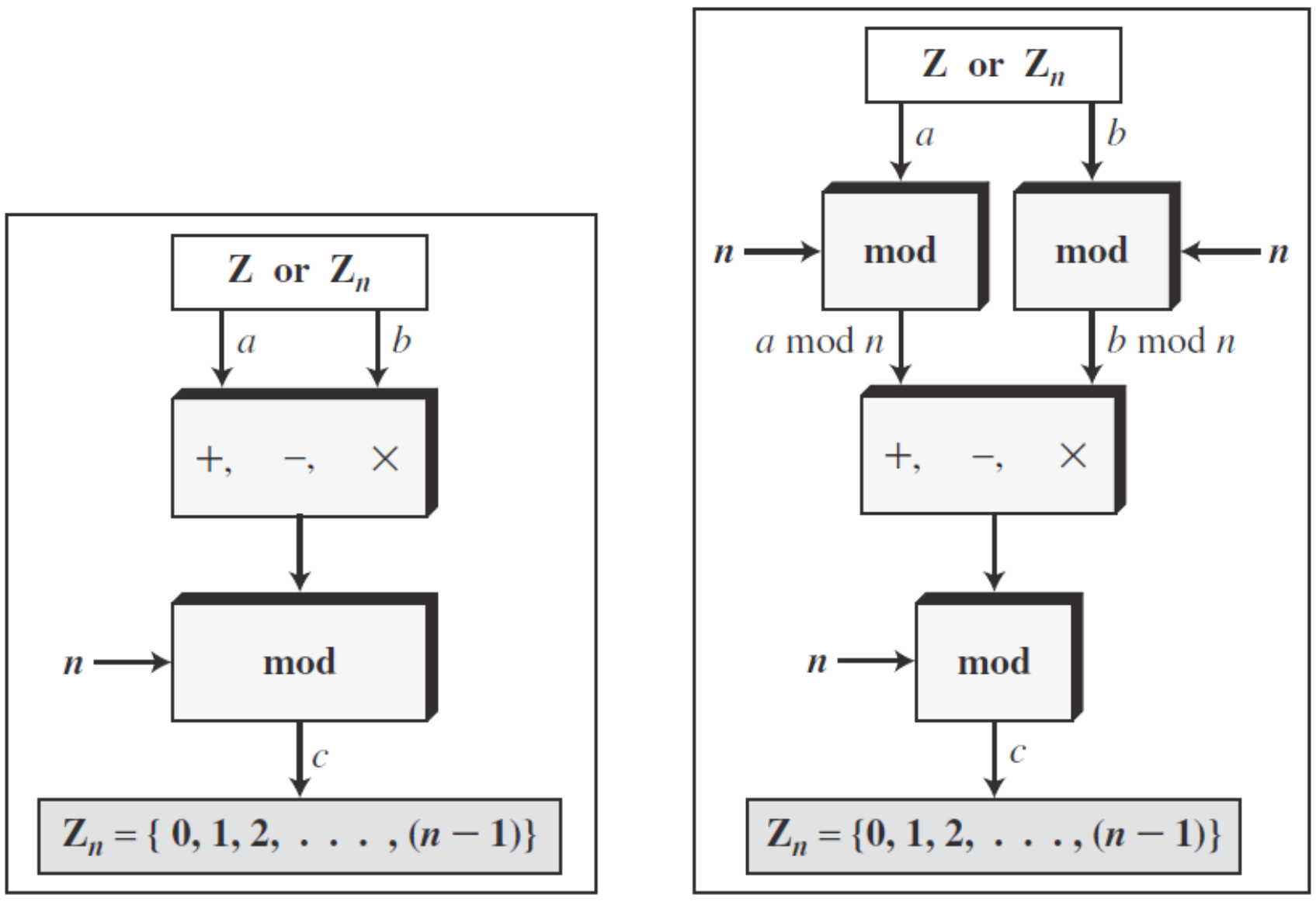

a. Original process    b. Applying properties

Fig.11: Properties of **mod** operator.

# III. BAMA

Compression catalyst begins selecting blocks from the binary sequence. The size of blocks depends on the size of binary sequence, however, it should never be less than 1024 bits, in fact, when the sequence binary is bigger (and therefore, blocks are larger) then, the performance of the final additional compression is better, automatically. Figure 12 shows the *ith* block of a binary sequence, where the subscripts *i* and *0* means, *ith* block and *0* treatment (i.e., original), respectivelly.

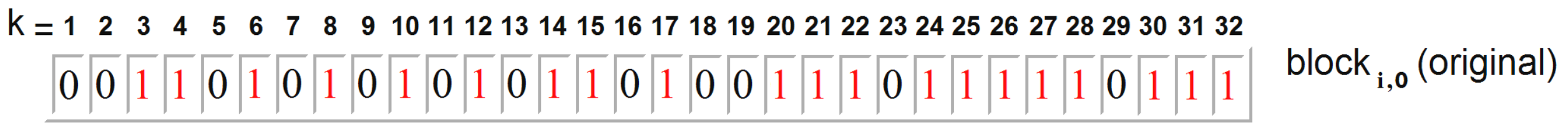

Fig.12: Block *ith* (original, i.e., without treatment) of the binary sequence.

BAMA has two modes of operation: Mode 1 (with ones subtraction) after each treatment, and Mode 2 (without ones subtraction) after each treatment. In the first one, the ones from each treatment were removed at the end of the same, however, in the second one, the ones of all treatments were removed at the end all together. The fundamental difference between them is that the second mode compresses more than first one, however, the cost of used memory increases. While, Fig.13 shows Mode 1, Fig.14 shows Mode 2.

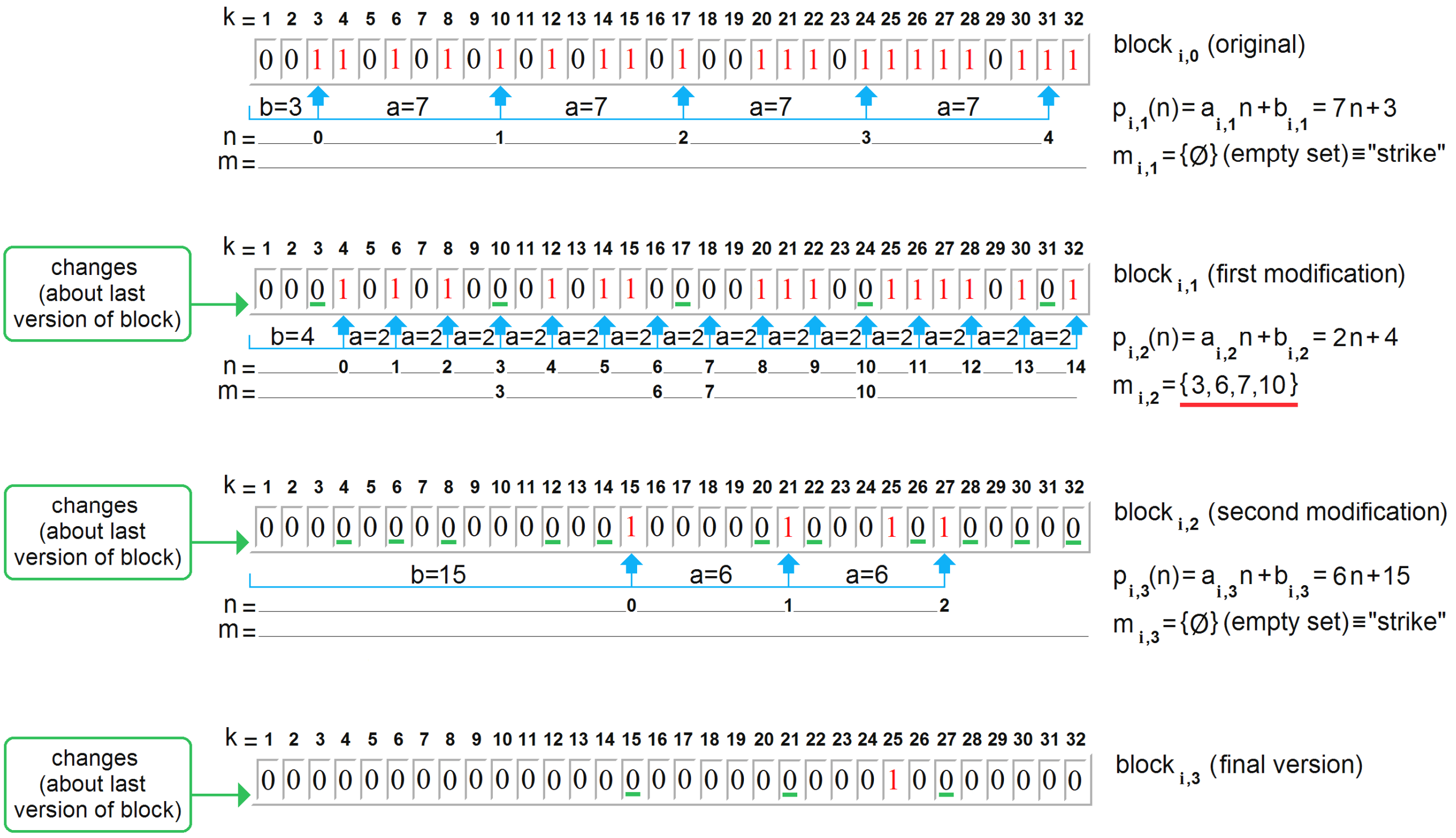

Fig.13: Mode 1.

**How they work?**

In Mode 1, top of Fig.13 shows the original block, i.e., $block_{i,0}$. The first treatment means a linear sweep on the block. To find the first 1, we set $b_{i,1}$, then we look for the second and third 1. If the distance between the third and second 1 is equal to the distance between the second and the first, apparently we found a $_{i,1}$, if not, we discard the second 1, and now we look for the fourth one. If the distance between the fourth and third 1 is equal to the distance between the third and the first, apparently we found a $_{i,1}$, if not, we discard the third 1, and so on. However, If we manage to involve several 1 inside the block, but sometimes fail (i.e., the 1 is a 0), the latter will call mutations. When, the mutation of a treatment is an empty set, this treatment is called "strike". In Fig.14 the mutation set is underscore in red. Besides, each $p_{i,j}$ is called "pike".

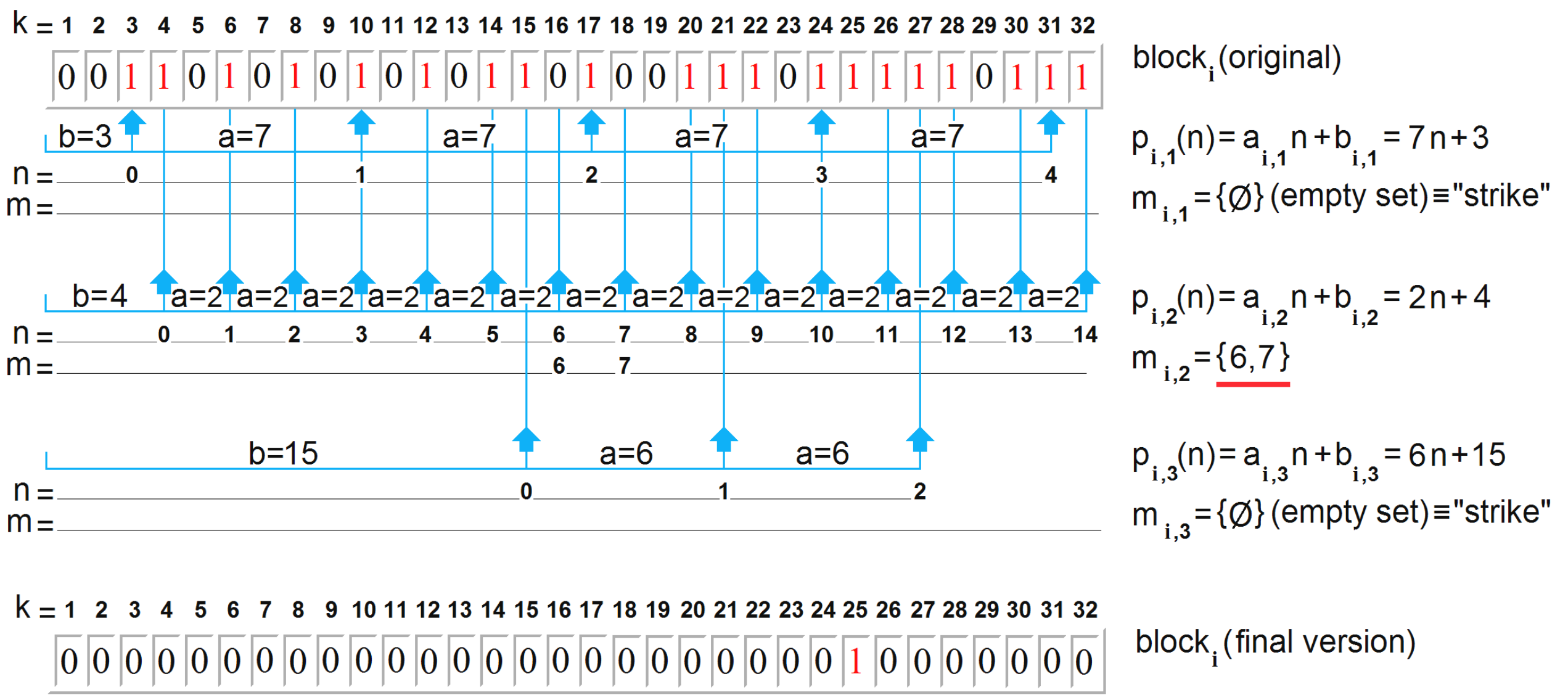

Fig.14: Mode 2.

From original to final block, we arrived through various treatments, in which successive triptongos $[a_{i,j}, b_{i,j}, m_{i,j}]$ are calculated. Although the frequency of mutations should preferably always low, the relationship between the amount of 1s taken at each pike, mutations and the block size constitute a delicate balance, whose goal is to increase the compression of the compressor significantly. The best estimate of the balance can be obtained via Genetic Algorithm, Linear Programming, etc.

A situation to be avoided at all costs is one where a pike contains to another, this spurious phenomenon is known as *hedging*. That is to say, $p_{i,j} \supset p_{i,k}\ \forall j \neq k$, because this causes a significant drop in the final compression performance.

On the other hand, Fig.13 shows the 0s suppressed in the above treatment underscore in green. The final result in both cases is the same (for this example, i.e., with a block of 32 bits, Fig.13 and 14), however, Mode 2 has minor quantity of mutations (2 vs. 4, Fig.13 vs. 14, respectively). In real examples (with block sizes bigger than Figures 13 and 14) the final results between both modes is substantially different, because the Mode 2 compresses more, without getting rid of the 1s the previous treatment to facilitate the formation of pikes (with fewer mutations). This inevitably increases the memory usage (buffers).

Finally, constitutes an important issue that BAMA does not wear windows as in the case of typical lossless compression algorithms [1]. Instead, it use blocks and warps for streaming applications (as we will mention in Section V). Besides, obviously this is a multipass procedure (multi treatment inside each block). However, it is also obvious that in the case of Mode 2, the blocks (and the treatments inside each block) may be processed in parallel or distributed.

## IV. Benchmarking Performance

Several measures are commonly used to express the performance of a compression method [1], i.e., the *compression performance*.

### *A. The Compression Ratio (CR) is defined as*

$$CR = \frac{size\ of\ the\ output\ stream}{size\ of\ the\ input\ stream} \tag{4}$$

A value of 0.6 means that the data occupies 60% of its original size after compression. Values greater than 1 imply an output stream bigger than the input stream (negative compression). The compression ratio can also be called bpb (bit per bit), since it equals the number of bits in the compressed stream needed, on average, to compress one bit in the input stream. In image compression, the same term, bpb stands for "bits per pixel." In modern, efficient text compression methods, it makes sense to talk about bpc (bits per character)—the number of bits it takes, on average, to compress one character in the input stream.

Two more terms should be mentioned in connection with the compression ratio. The term *bitrate* (or "bit rate") is a general term for bpb and bpc. Thus, the main goal of data compression is to represent any given data at low bit rates. The term *bit budget* refers to the functions of the individual bits in the compressed stream. Imagine a compressed stream where 90% of the bits are variable-size codes of certain symbols, and the remaining 10% are used to encode certain tables. The bit budget for the tables is 10%.

### *B. Compression Factor (CF)*

$$CF = \frac{size\ of\ the\ input\ stream}{size\ of\ the\ output\ stream} \tag{5}$$

That is to say, it is the inverse of the CR. In this case, values greater than 1 indicate compression and values less than 1 imply expansion. This measure seems natural to many people, since the bigger the factor, the better the compression. This measure is distantly related to the sparseness ratio, a performance measure discussed in [1].

### *C. Compression Performance (CP)*

$$CP = 100 \times (1 - CR) \tag{6}$$

This expression is also a reasonable measure of compression efficiency. A value of 60 means that the output stream occupies 40% of its original size (or that the compression has resulted in savings of 60%).

### *D. Bits per Pixel (bpp)*

In image compression, the quantity bpp (bits per pixel) is commonly used. It equals the number of bits needed, on average, to compress one pixel of the image. This quantity should always be compared with the bpp before compression.

### *E. Compression Gain (CG)*

$$CG = 100 \log_e \left( \frac{reference\ size}{compressed\ size} \right) \tag{7}$$

where the reference size is either the size of the input stream or the size of the compressed stream produced by some standard lossless compression method. For small numbers $x$, it is true that $\log_e(1 + x) \approx x$, so a small change in a small compression gain is very similar to the same change in the compression ratio. Because of the use of the logarithm, two compression gains can be compared simply by subtracting them. The unit of the compression gain is called *percent log ratio* and is denoted by $CG = 100 \times \left( \frac{reference\ size}{compressed\ size} - 1 \right)$.

*F. Cycles per Byte (CPB)*

The speed of compression can be measured in *cycles per byte* (CPB). This is the average number of machine cycles it takes to compress one byte. This measure is important when compression is done by special hardware.

*G. Compression Catalyst Factor per Block (CCRPB)*

$$CCFPB_i = \frac{entropy\ coding\ of\left(original\ block_i\right)}{entropy\ coding\ of\left(final\ block_i \cup [a_{i,j}, b_{i,j}, m_{i,j}]\right)} \quad \forall i, j \tag{8}$$

where $\cup$ means "*union*", subscript $i$ represents the *ith* block, and $j$ the *jth* triphthong $[a_{i,j}, b_{i,j}, m_{i,j}]$.

This is one of the two primary metrics for the quality assessment of compression in this paper, because, it reveals like no other input of the catalyst in the compression process of each block. Its value is in the following range,

$$1.2 \le CCFPB_i \le 2 \tag{9}$$

The other is the mean *CCFPB* (or $CCFPB_{mean}$), which is the most complete estimation of the real contribution of the catalyst to total compression process. Such new metric is represented by

$$CCFPB_{mean} = \frac{1}{NB} \sum_{i=1}^{NB} CCFPB_i \tag{10}$$

where $CCFPB_{mean}$ is the mean *CCFPB*, which in most cases coincides with the individuals $CCFPB_i$, being *NB* the number of blocks.

In this work, we have generated several binary sequences of $2^{20}$ elements thanks to MATLAB® [28] language, with different size of blocks. Fig.15 shows the employed comparison scheme, with and without compression catalyst (BAMA). The lossless compression algorithm used in first place was Huffman [1]. Table I

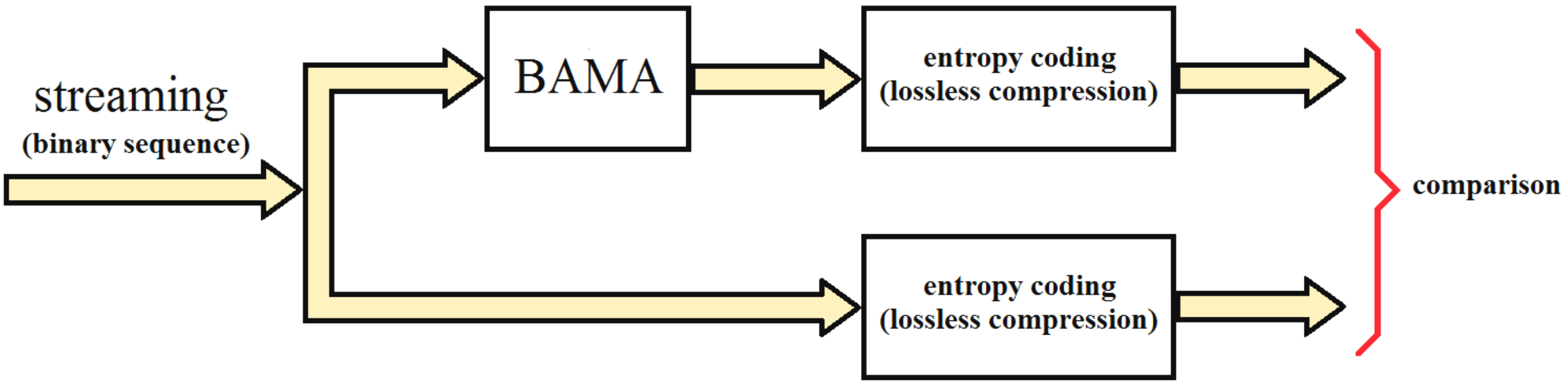

Fig.15: Comparison scheme with and without compression catalyst.

TABLE I

METRICS VS. HUFFMAN, WITH AND WITHOUT BAMA (MODES 1 AND 2)

| METRICS | HUFFMAN | BAMA (MODE 1) & HUFFMAN | BAMA (MODE 2) & HUFFMAN |
|---|---|---|---|
| CR | 0.1652 | 0.1032 | 0.1001 |
| CF | 6.0530 | 9.6848 | 9.9874 |
| CP | 83.48 | 89.68 | 89.99 |
| CG | 78.19 | 98.60 | 99.94 |
| $CCFPB_{MEAN}$ | - | 1.6008 | 1.6501 |

shows the metrics vs Huffman, with and without BAMA (Modes 1 and 2). As we can see on Table I, $CCFPB_{mean}$ is 1.6008 for Mode 1 and 1.6501 for Mode 2, which meets the above. Besides, Table I shows that BAMA+ Huffman have 60 % (for Mode 1) and 65 % (for Mode 2) more compression than Huffman alone, which amply justifies the use of the catalyst.

TABLE II

METRICS VS. ARITHMETIC, WITH AND WITHOUT BAMA (MODES 1 AND 2)

| METRICS | ARITHMETIC | BAMA (MODE 1) & ARITHMETIC | BAMA (MODE 2) & ARITHMETIC |
|---|---|---|---|
| CR | 0.1675 | 0.1060 | 0.1021 |
| CF | 5.9700 | 9.4326 | 9.7908 |
| CP | 83.25 | 89.40 | 89.79 |
| CG | 77.59 | 97.46 | 99.08 |
| $CCFPB_{MEAN}$ | - | 1.5879 | 1.6011 |

Table II shows the metrics vs Arithmetic, with and without BAMA (Modes 1 and 2). As we can see on Table II, $CCFPB_{mean}$ is 1.5879 for Mode 1 and 1.6011 for Mode 2, which meets the above. Besides, Table II shows that BAMA+ Arithmetic have 58 % (for Mode 1) and 64 % (for Mode 2) more compression than Arithmetic alone, which amply justifies the use of the catalyst.

Finally, Fig. 16 shows us the relationship between the Probability of finding the next 1 vs the frequency of occurrence of the same ones. When we desire a high frequency of occurrence of the 1s that meet the modular condition, then their frequency falls inversely related. This is obvious, since if we intend to find fewer 1s, this is more probably to happen.

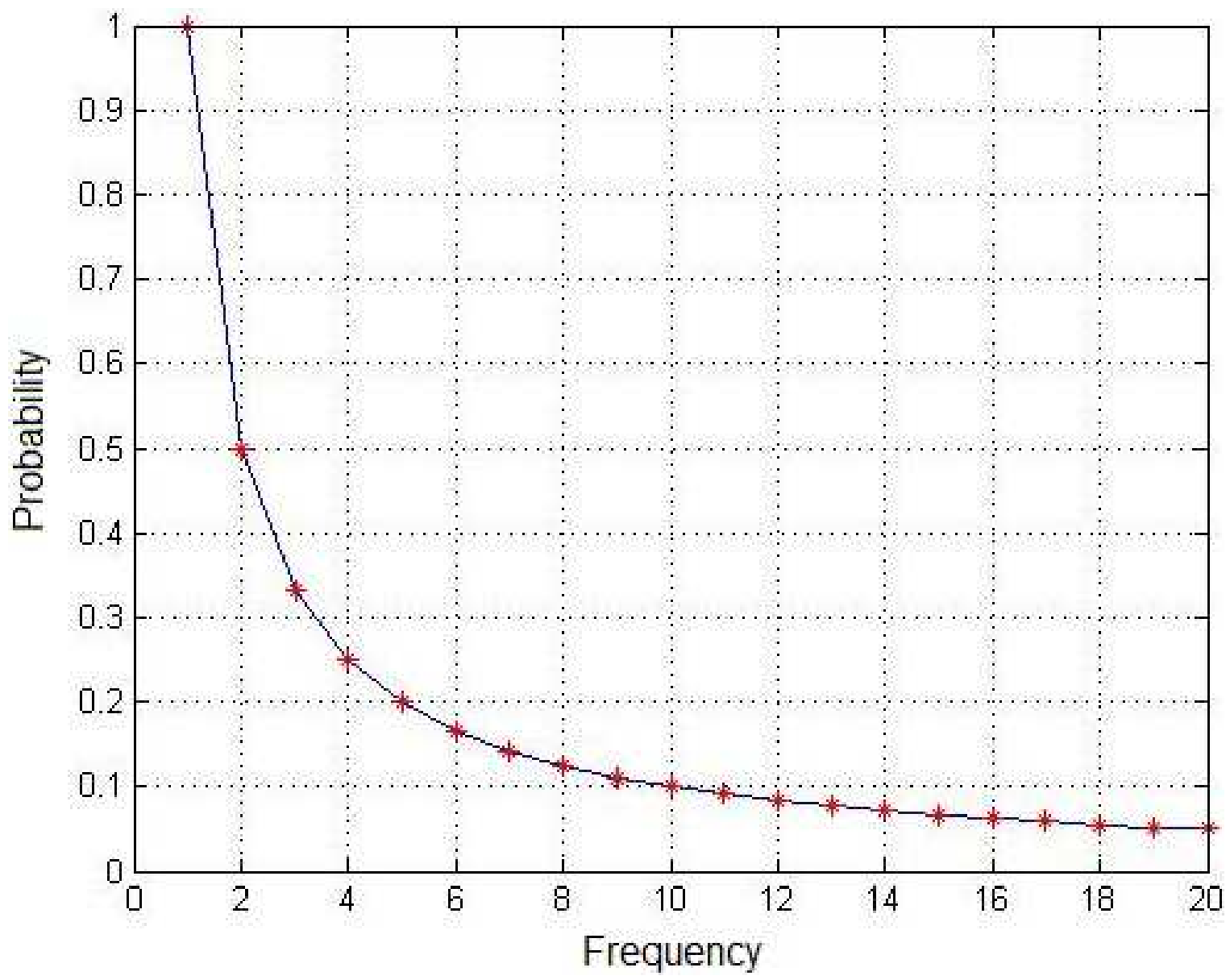

Fig.16: Probability vs frequency.

## V. APPLICATIONS

The main applications of BAMA are: additional storage for Big Data systems, data base of Bioinformatics sequences, human-computer interaction, digital documents, digital humanities, among others.

On the other hand, BAMA is emerging as particularly suitable to be applied to Digital Image Processing [43], Digital Audio [44-48], Digital Video [49-51], Audio Streaming [52-54], and Video Streaming [55-58]. Specifically, entropy coding sector inside, previous to lossless compression algorithm [1-5].

Finally, and although we have seen in this paper to BAMA applied to binary sequences, the same goes for other types of sequences non binaries, that is to say, all type of sequences (for all types of characters). It is important to note here that although we have seen in this paper to BAMA applied on a linear or unidimensional array (binary sequence), it can be applied to multidimensional arrays, as is the case of the lossless part of codecs for TV Digital [6, 7], Digital Video [49-51], and Video Streaming [55-58].

## VI. CONCLUSIONS

This paper presented a novel concept in encoding schemes for lossless compression known as compression catalyst, and called BAMA. Such catalyst does not compress, but facilitates the action of the real compressor. In this paper, they were Huffman and Arithmetic (however, similar results can be obtained with RLE and LZW), that is, BAMA acts as a compression catalyst. Finally, this catalyst allows a significant increase in the compression performance of binary sequences, among others.

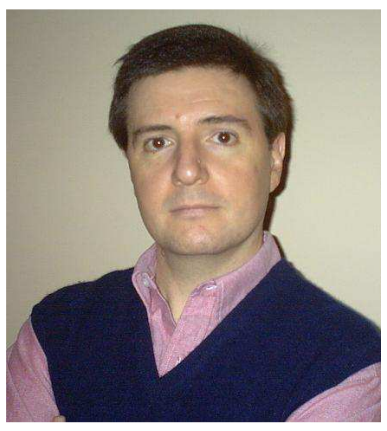

**Mario Mastriani** received the B.Eng. degree in 1989 and the Ph.D. degree in 2006, both in electrical engineering. Besides, he received the Ph.D. degree in Computer Sciences in 2009 and the Ph.D. degree in Science and Technology in 2011. He was the director of research and development laboratories in signal and image processing, technical innovation, and new technologies in several institutions. Currently, he is a CoFounder and the Chief Science Officer (CSO) of DLQS LLC, 4431 NW 63RD Drive, Coconut Creek, FL 33073, USA. He published 45 papers. He was a reviewer of IEEE Transactions on Neural Networks, Signal Processing Letters, Transactions on Image Processing, Transactions on Signal Processing, Communications Letters, Transactions on Geoscience and Remote Sensing, Transactions on Medical Imaging, Transactions on Biomedical Engineering, Transactions on Fuzzy Systems, Transactions on Multimedia; Springer-Verlag Journal of Digital Imaging, SPIE Optical Engineering Journal; and Taylor & Francis International Journal of Remote Sensing. He was a member of IEEE, Piscataway, USA, during 9 years. He became a member of WASET in 2004. His areas of interest include Quantum Signal and Image Processing, and cutting-edge technologies for computing and communication.